\documentclass{emulateapj}
\slugcomment{{\sc Accepted to ApJ:} October 13, 2008} 
\usepackage{amsmath}
\usepackage{amssymb}
\usepackage{graphicx}

\usepackage{natbib}
\newcommand{\lSect}[1]{{\label{sec:#1}}}

\def\gtaprx {\lower .1ex\hbox{\rlap{\raise .6ex\hbox{\hskip .3ex
	{\ifmmode{\scriptscriptstyle >}\else
		{$\scriptscriptstyle >$}\fi}}}
	\kern -.4ex{\ifmmode{\scriptscriptstyle \sim}\else
		{$\scriptscriptstyle\sim$}\fi}}}
\def\ltaprx {\lower .1ex\hbox{\rlap{\raise .6ex\hbox{\hskip .3ex
	{\ifmmode{\scriptscriptstyle <}\else
		{$\scriptscriptstyle <$}\fi}}}
	\kern -.4ex{\ifmmode{\scriptscriptstyle \sim}\else
		{$\scriptscriptstyle\sim$}\fi}}}
\newcommand{\note}[1]{\emph{\textcolor{red}{}}}
\newcommand{\code}[1]{{\tt{#1}}}
\newcommand{\Msun}{{\ensuremath{\mathrm{M}_{\odot}}}}

\newcommand{\Ni}{{\ensuremath{^{56}\mathrm{Ni}}}}

\newcommand{\Co}{{\ensuremath{^{56}\mathrm{Co}}}}
\newcommand{\He}{{\ensuremath{^{4} \mathrm{He}}}}
\newcommand{\Hy}{{\ensuremath{^{1} \mathrm{H}}} }
\newcommand{\Ox}{{\ensuremath{^{16}\mathrm{O}}} }
\newcommand{\Ti}{{\ensuremath{^{44}\mathrm{Ti}}}}
\newcommand{\Si}{{\ensuremath{^{28}\mathrm{Si}}}}

\newcommand{\Sectff}[1]{{\ref{sec:#1}}}
\newcommand{\Sect}[1]{{\S~\Sectff{#1}}}

\begin{document}

\title{Mixing in Zero and Solar Metallicity Supernovae}

  \author{C.C.
  Joggerst\altaffilmark{1,}\altaffilmark{2}, S. E. Woosley\altaffilmark{1},
  \& Alexander Heger\altaffilmark{2,}\altaffilmark{3}}

\altaffiltext{1}{Department of Astronomy and Astrophysics,
University of California at Santa Cruz, Santa Cruz, CA 
95060; cchurch@ucolick.org}

\altaffiltext{2}{Theoretical Astrophysics (T-6), Los Alamos National
Laboratory, Los Alamos, NM 87545}

\altaffiltext{3}{School of Physics and Astronomy, University of
  Minnesota,Minneapolis, MN 55455}

\begin{abstract}
Two-dimensional simulations of mixing and fall back in
non-rotating massive stars have been carried out using realistic
initial models for the presupernova star and assuming standard
spherically symmetric explosions of $1.2 \times 10^{51}$erg.  Stars of
15 and 25 M$_{\sun}$ with both primordial and solar composition were
modeled.  The zero metallicity supernova progenitors were compact blue
stars and the amount of Rayleigh-Taylor induced mixing in them was
greatly reduced compared with what was seen in the red supergiants
with solar metallicity.  The compact zero-metal stars also experienced
more fallback than their solar metallicity counterparts.  As a result,
the ejected nucleosynthesis from the two populations was very
different.  For the simple explosion model assumed, low metallicity
stars ejected ejected too little iron and intermediate mass elements
even to explain the abundance patterns in the most iron-poor stars
found to date, suggesting that some important ingredient is missing.
Rotation is likely to alter these conclusions by producing a greater
fraction of red supergiants among Population III stars.  The
velocities of the heavy elements in all models considered - both red
and blue supergiants - were less than observed in SN 1987A,
suggesting that at least occasionally, asymmetric aspects of the
explosion mechanism and fallback play a major role in mixing.
\end{abstract}


\maketitle

\section{INTRODUCTION}
\lSect{introduction}

The nucleosynthetic yields of supernovae are important components of
galactic chemical evolution and are essential to understanding the
abundances observed in metal-poor stars.  The yield of a core-collapse
supernova is determined by its presupernova evolution, the geometry
and energy of its explosion, and by the mixing and fallback that occur
as the supernova shock traverses the star \citep[e.g.,][]{Chevalier:2005}.
Mixing and fallback also affect the supernova's light curve and
spectrum and the appearance of its remnant.  The dense knots and
filaments visible in Hubble \citep{Blair:2000} and Chandra \citep{Hughes:2000}
images of supernova remnants provide clear evidence that mixing of
some sort is a common occurrence.  Supernova 1987A, the closest and
most thoroughly observed modern supernova, provided important
observational constraints on models.  Its smooth bolometric light
curve required extensive mixing of the helium core with the hydrogen
envelope \citep[e.g.,][]{Woosley:1988}.  The early appearance of X-rays and
$\gamma$-rays from radioactive decay of $^{56}$Ni, as well as spectroscopic
evidence some fraction of the iron peak was mixed out to ~$4000$ km
s$^{-1}$, provided further evidence for extensive mixing in the
interior of the supernova at early times \citep{Arnett:1989,Witteborn:1989,Utrobin:2004}.

Differences in the presupernova structure of zero and solar
metallicity stars alter the way mixing and fallback processes
operate. Below about 40 \Msun, non-rotating, zero-metallicity stars
are expected to be more compact than solar-metallicity progenitors
\citep{Heger&Woosley:2008,Hirschi:2008}.  Metal-free gas has a lower opacity than
solar-metallicity gas and lacks initial seed nuclei for the CNO cycle,
leading to inefficient hydrogen burning and a very dense hydrogen
shell of low entropy in the presupernova stars.  \citet{Chevalier:1989}
predicted that higher amounts of fallback are expected for more
compact progenitors. In a recent paper, \citet{Zhang:2008} used their
one-dimensional Eulerian code \code{PANGU} to determine the remnant
masses left behind by the supernova models calculated in surveys by
\citet{Woosley&Heger:2007} and \citet{Heger&Woosley:2008}.  They found that zero-metallicity
supernovae experienced more fallback and left behind larger compact
remnants than their solar metallicity counterparts.  For example, the
baryonic remnant masses left behind by 25 \Msun \ stars of zero and
solar metallicity were 4.16 and 2.09 \Msun, respectively.

Different presupernova structures, arising from differences in stellar
mass and metallicity, determine where Rayleigh-Taylor instabilities
occur and the extent to which they grow.  An initially static,
incompressible fluid is unstable if the pressure gradient points
opposite to the density gradient, i.e., when $(dP/dr)(d\rho/dr) < 0$
\citep[e.g.][]{Chevalier:1976,Benz:1990}.  The location of these density inversions
varies with time as the forward and reverse shocks propagate through
the star.  Particularly important are regions where the forward shock
encounters an increasing value for $\rho r^3$, where $\rho$ is the
density and $r$, the radius \citep{Herant&Woosley:1994}.  The time scale also
depends upon the initial stellar structure.  In particular, a more
compact star will experience faster shock propagation, leaving less
time for instabilities to grow.

Because SN 1987A was so well observed, most previous studies of
Rayleigh-Taylor mixing in core collapse supernovae
\citep{Arnett:1989,Fryxell:1991,Muller:1991,Hachisu:1990,Hachisu:1992,Herant:1991,Herant&Benz:1992,Kifonidis:2006} have been in
the context of that event.  Others have studied red supergiant
progenitors \citep{Herant&Woosley:1994}.  Except for Herant and Woosley, these
studies all used red and blue supergiants of 15 to 20 \Msun \ as
progenitors. The methodology of all these studies was similar. A
one-dimensional progenitor model was exploded, somewhat artificially,
by means of a piston or a thermal bomb, and the subsequent evolution
followed with a two-dimensional code.

More recently, \citet{Kifonidis:2003} and \citet{Kifonidis:2006} have used a different
approach. These authors followed a blue supergiant model from the
first seconds of the explosion out to about 5 days after core
collapse, using first one code with neutrino physics for the early
times, and another code with mesh refinement for later times.  They
saw mixing at the Si-O interface, a location at which no previous
studies had found mixing, and were able to reproduce the high
$^{56}$Ni velocities observed in SN 1987A, something previous studies
had not done.

While attempts to reproduce observations of 1987A have been numerous,
no multidimensional studies of mixing and fallback in very low
metallicity supernovae have been done.  The nucleosynthetic yields of
metal-free (Pop III) and extremely metal-poor (EMP) stars might still
be visible in the abundance patterns observed in some halo stars in
our own galaxy. Of particular interest are the ``ultra-iron-poor''
(HMP) stars \citep{Frebel:2005,Aoki:2006}.  These stars with [Fe/H]$<$ -5, have
abundance patterns that differ considerably from those observed in
stars with near solar metallicity or even other metal-poor stars
\citep{Cayrel:2004}. It is possible that these iron-poor stars were enriched
by only one or a few supernovae \citep{Frebel:2005}.  In particular, the two
most metal-poor stars known and several other UMP stars display marked
enhancement in C, N, and O relative to Fe.  Previous studies
\citep{Iwamoto:2005,Tominaga:2007,Heger&Woosley:2008} have sought to explain these abundance
patterns with one-dimensional models for supernovae that parametrize
the amount of mixing and fallback to match what is
observed. Simulating mixing and fallback directly, rather than
parametrically, requires a multi-dimensional approach.

In this paper, we use two-dimensional axisymmetric simulations to
explore directly the amount of Rayleigh-Taylor-induced mixing that
occurs in non-rotating zero- and solar- metallicity stars.  Our
methodology is similar to the earlier studies (before Kifonidis) of SN
1987A. In \Sect{methodology}, we discuss our initial models, our
modifications to the \code{FLASH} code, and our simulation setup.  In
\Sect{results}, results are given that show the degree of mixing, the
final velocity distribution of isotopes, and the ejected yields.
These yields are compared with abundances observed in HMP stars in
\Sect{yields}.  Finally, we provide a short summary of results
and their interpretation in \Sect{conclusions}

\section{MODELS AND METHODS}
\lSect{methodology}

The present work follows the method used in many previous studies of
mixing in supernovae. A one-dimensional code was used to evolve and
explode the pre-supernova model and to follow the first stages of the
expansion to the time when the reverse shock was just beginning to
form.  No significan growth of instabilities is expected before the
formation of the reverse shock.  The one-dimensional model was then
mapped onto a two-dimensional grid and the ensuing instabilities
followed.

While \citet{Kifonidis:2006} reproduced the observations of SN 1987A somewhat
better than previous attempts, possibly by following the early stages
of the explosion, this paper does not do that.  The physics of the
initial explosion remains uncertain and our goal is to isolate the
differences in post-explosive mixing that arise as a direct
consequence of the differences in initial structure of the
pre-supernova models. Exploding the star with a piston in the same
location and with the same energy in all models allows us to
accomplish this.

The choice of piston mass location is constrained by observational
parameters.  The piston cannot be located within the iron core or the
resulting explosion will produce far too much of 54,58 Fe and other
neutron-rich species to be in agreement with observations of these
isotopes.  On the other hand, the remnant mass will be too large to
agree with observations if the explosion site is located outside the
base of the oxygen shell.  There are reasons to believe the location
site is located at the base of the oxygen shell--the large density
decrease associated with this location is dynamically important, and
successful explosion calculations often find the mass cut there.

The papers from which our models are taken also reported on models in
which the stars were exploded with a piston at the edge of the ni
core.  These explosions experience slightly less fallback, and produce
more nickel, by a factor of 2 or so, than the models presented in this
paper.

\subsection{Progenitor Models}
\lSect{kepler}

Initial models were taken from the surveys of \citet{Heger&Woosley:2008} and
\citet{Woosley&Heger:2007}.  Both of these papers used the \code{KEPLER} code
\citep{Weaver:1978,Woosley:2002} to evolve stars through all stable stages of
nuclear burning until their iron cores became unstable to collapse.
At this point, pistons located at or near the base of the oxygen shell
were used to explode the stars.  \citet{Heger&Woosley:2008} simulated the evolution
and explosion of 10 to 100 \Msun \ stars with zero initial
metallicity.  Explosion energies ranged from 0.3 to 10 B, where 1
Bethe = 1 B = $10^{51}$ ergs.  \citet{Woosley&Heger:2007} examined solar-metallicity
stars from 12 to 100 \Msun \ which were exploded by pistons similar to
the other survey, but for a more limited set of masses and energies.
Both surveys were limited to non-rotating progenitors.  The
zero-metallicity stars were assumed to have no mass loss, while the
solar-metallicity models took mass loss into account.

Here only two representative stars from each survey are studied:
Models z15D and z25D from \citet{Heger&Woosley:2008} and Models s15A and s25A from
\citet{Woosley&Heger:2007}.  The letter ``z'' indicates zero initial metallicity,
while ``s'' indicates solar metallicity.  The numbers in the models
correspond to the initial mass of the star in \Msun \ and the final
letter is the explosion energy, 1.2 B in each case.  The piston was
located at the place in the star where the entropy was equal to $4.0
k_B/$baryon. This corresponded to the base of the oxygen shell. Series
sA and zD are thus directly comparable in all respects save their
initial metallicity.  15 and 25 \Msun \ represent the ``canonical''
supernova cases with the most commonly employed explosion energy and
piston location.  15 and 25 \Msun \  stars are also in the same mass
range as previous studies of Rayleigh-Taylor mixing in supernovae,
allowing for easy comparison.

The one-dimensional progenitor models used in this study lead to
spherically symmetric explosions.  Mapping the models from one to two
dimensions after the explosion has taken place has the effect of
suppressing low-order departures from spherical symmetry.  This work
only addresses spherically symmetric explosions with asymmetries of
significantly higher mode than $l=1$ or 2.

\subsection{The FLASH code}
\lSect{FLASH}

The \code{FLASH} code \citep{Fryxell:2000} was used to follow shock
propagation and mixing in the models in two dimensions. This code has
been extensively verified and tested \citep{Calder:2002,Weirs:2005}.
\code{FLASH} is an adaptive mesh refinement code based on an Eulerian
implementation of the Piecewise Parabolic Method (PPM) of
\citet{Colella&Woodward:1984}. The code can be configured in a number of different
ways.  We used the HLLE Riemann solver to resolve shocks.
\code{KEPLER} uses 19 different isotopes to evolve stellar models
through stable stages of nuclear burning.  We use the ``aprox19''
composition module included with the \code{FLASH}2.5 distribution to
map these isotopes directly to \code{FLASH}.  The abundances of 19
different isotopes, from $^1$H to $^{56}$Ni, are stored for each grid
cell.  Explosive nuclear burning could have been followed in
\code{FLASH} using this network, but the burning was over by the time
the models were mapped into \code{FLASH}.  Mapping the star in at
earlier times to follow the explosive burning in two dimensions would
have had little effect, as no departure from spherical symmetry is
expected until the formation of the reverse shock, which occurs much
later in the calculation.  \code{FLASH} was configured to use
axisymmetric coordinates.  The gravitational potential was computed
using a multipole method to solve Poisson's equation.  The mass
distribution in this simulation had only slight deviations from
spherical symmetry, so the additional gravitational force from
overdensities in the simulations was small.  The gravitational
potential from the point mass at the origin of the grid was added to
the potential computed by the multipole solver at each time step.  The
hydrodynamic equations were solved using an explicit,
dimensionally-split approach: the hydrodynamic equations were solved
first along one coordinate grid direction, then the other at each time
step.

An equation of state was employed that assumed full ionization and
included contributions from radiation and ideal gas pressure:

\begin{equation}
        P= \frac{1}{3} a T^4 + \frac{k_{B} T \rho}{m_p \mu}
\end{equation}
\begin{equation}
       E = \frac{a T^4}{\rho} + 1.5\frac{k_{B} T}{m_p \mu}
\end{equation}

where $P$ is the pressure, $a$ is the radiation constant, $k_{B}$ is
Boltzmann's constant, $T$ is the temperature, $\rho$ is the density,
$m_p$ is the proton mass, $\mu$ is the mean molecular weight, and $E$
is the energy.  Although the outer regions of the remnant may not always
be fully ionized, the regions where Rayleigh-Taylor mixing takes place
are.  Exploratory simulations performed in one dimension with
\code{FLASH} using a Helmholtz equation of state were identical to
one-dimensional simulations performed with the perfect gas and
radiation equation of state used in the simulations presented in this
paper.  These one-dimensional simulations were only run to $2 \times
10^4$ seconds.  At later times, the density is low enough that the
Helmholtz equation of state no longer applies.

The code was configured to use pressure, density, $^4$He, and $^{16}$O
as its refinement variables.  An error estimate for a block was
computed using the second derivative of the chosen refinement
variables.  If the estimated normalized error in one or more of these
variables was greater than a given value, regions were refined until a
normalized error is reached that is less than the acceptable value or
the maximum level of refinement is reached.  Regions of the
simulations with steep gradients in one or more of these variables
were likely to be refined. If the normalized error estimate was below
a certain value, i.e., the absolute value of the second derivative of
one or more variables was small, then the region was ``de-refined''.

\subsection{Modifications to the \code{FLASH} Code}
\lSect{modifications}

The \code{FLASH}2.5 release was customized to include a module that
inserts a roughly circular zero-gradient inner boundary around the
origin.  This prevented infalling matter near the origin in the
simulations from backing up and affecting the outer regions.  Matter
was allowed to fall though the zero-gradient, quasi-circular boundary
at the center of the model and accumulate on the point mass at the
origin. If the radius for the inner boundary passed through one of the
inner zones, the zones interior to that zone were set to be duplicates
of the zone on the boundary.  While the Cartesian nature of the
axisymmetric coordinate system meant that the inner boundary was only
as close to circular as one can reproduce with square components, it
did not introduce a significant amount of error into the calculation.
The inner boundaries were chosen to be within the sonic radius,
ensuring that small numerical errors at the interior boundary did not
accumulate and affect the flow of fluid upstream from the
boundary. The sonic point moved outward, not inward, as the simulation
time progressed, so the inner boundary was always within the sonic
radius.  The falling temperature caused the sound speed to decline
with time, while the velocity near the inner boundary increased with
time.  The gravitational potential resulting from this point mass was
updated and added to the gravitational forces computed by the
multipole solver at each time step.

A module was also included that locally deposited energy from
radioactive decay of $^{56}$Ni to $^{56}$Fe. The amount of energy
\begin{equation}
               dE_{\Ni} = \lambda_{\Ni} X_{\Ni}
               e^{-\lambda_{\Ni}*t}*q(\Ni)
\end{equation}
The decay rate of \Ni, $\lambda_{\Ni}$, is $1.315\times10^{-6}$
s$^{-1}$, and the amount of energy released per gram of decaying \Ni \
is $q(\Ni)$, for which we took the value $2.96 \times 10^{16}$
erg g$^{-1}$.  $X_{\Ni}$ is the fraction of \Ni \  in the block.  The
amount of \Co \ at a given time could be found as a function of the
amount of initial \Ni \  by
\begin{equation}
               X_{\Co} =
               \frac{\lambda_{\Ni}}{\lambda_{\Co}-\lambda_{\Ni}}
               X_{\Ni} (e^{-\lambda_{\Ni}t}-e^{-\lambda_{\Co}*t})
\end{equation}
so that the energy deposition rate from \Co \ as a function of time was
given by
\begin{equation}
               dE_{\Co} =
               \frac{\lambda_{\Ni}}{\lambda_{\Co}-\lambda_{\Ni}}
               X_{\Ni} (e^{-\lambda_{\Ni} t}-e^{-\lambda_{\Co} t}))
               \lambda_{\Co}*q(\Co)
\end{equation}
We assumed a decay rate for \Co, $\lambda_{\Co}$, of $1.042
\times10^{-7}$ s$^{-1}$, and an energy per gram of decaying \Co,
 $q(\Co)$, for which we took the value of  $6.4 \times 10^{16}$
erg g$^{-1}$. 

\subsection{Calculations}
\lSect{calculations}

The supernova models were evolved with \code{KEPLER} to the point
where all explosive nuclear burning had ceased and the reverse shock
had just begun to form. This occurred at $10^3$ s and $10^4$ s for
15 and 25 \Msun \ stars of solar composition, and at $25$ s and
$100$ s for 15 and 25 \Msun \ stars of primordial composition,
respectively.  At these times, the one dimensional models from
\code{KEPLER} were mapped onto a two-dimensional axisymmetric grid and
evolved forward in time with the \code{FLASH} code. A similar
simulation of Model s25A was also performed using a progenitor evolved
to $10^3$ seconds with \code{KEPLER} before being mapped to
\code{FLASH}.  No significant difference in the later evolution of
s25A models evolved to $10^3$ seconds and $10^4$ seconds was observed.
Only one quadrant of the star was carried in the calculation,
enforcing symmetry about the left $y$- and bottom $x$-axes, while
allowing material to leave the grid through a zero-gradient boundary
at the right $y$- and top $x$-axes.

An enhanced flow was observed along the $x$- and $y$-axes.  This flow
was not large in comparison with the rest of the simulation, but it
was present, as can be seen in Figures \ref{s15A_dens} -
\ref{z25D_dens}.  This is a well-documented artifact of the
dimensionally-split approach to solving the hydrodynamic equations.
It did not substantially influence the evolution of the simulations
presented here.

\begin{figure*}
  \includegraphics[width=0.3\textwidth]{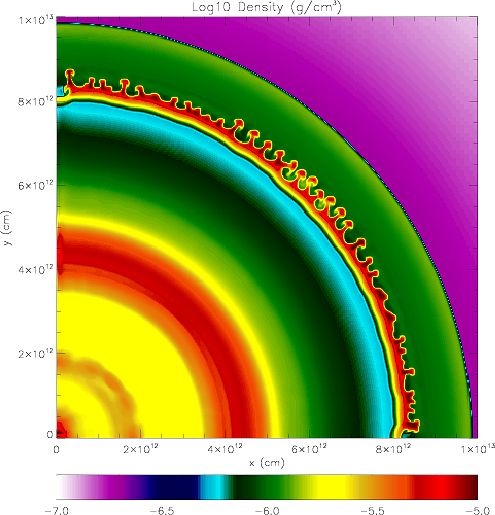}
  \includegraphics[width=0.3\textwidth]{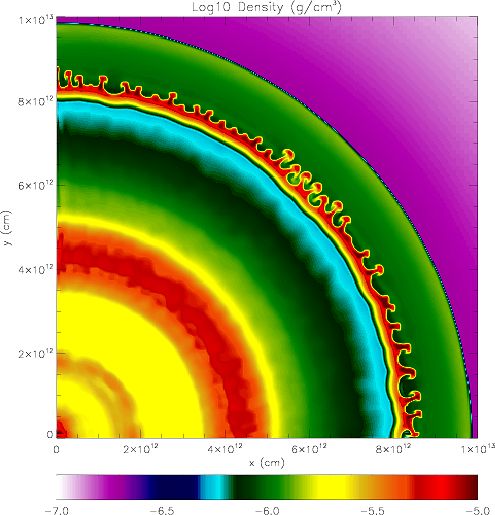}
  \includegraphics[width=0.3\textwidth]{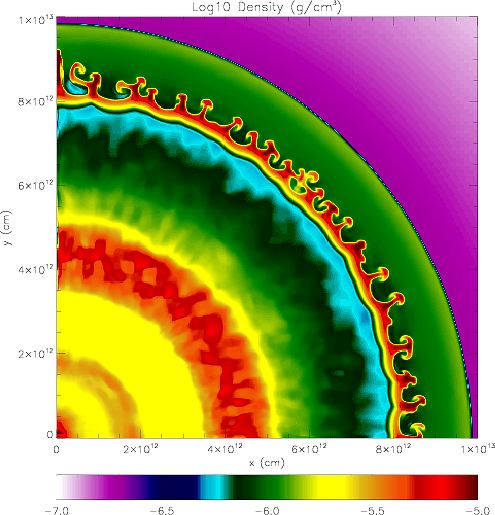}
  \caption{Influence of perturbations on the initial scale of the
    instability.  Panel (a): no additional perturbation.  Panel (b):
    random perturbations of $0.5\%$ applied to velocity.  Panel (c):
    random perturbations of $2.0\%$ applied to velocity. Panel (a) and
    (b) show instability growth on nearly the same scale, implying
    that grid perturbations were on the order of $0.5\%$.  The scale
    of the instability in panel (c) is noticeably larger, showing that
    perturbations from the grid are no larger than $2.0\%$.  The other
    models presented in the current paper, not shown, are nearly the
    same.  The initial scale and spectrum of the instability is washed
    out after the instability becomes nonlinear in the solar
    stars.\label{s15A_pert}}
\end{figure*}

Perturbations arising from a Cartesian grid are also inevitable.  In
order to quantify these grid effects, we performed simulations of all
stars with a random perturbation in velocity with a maximum amplitude
of $0.5\%$ and $2.0\%$.  We also performed simulations with no
additional perturbation. We found that perturbations of $2.0\%$ in
velocity had a clear effect on the initial scale of the
Rayleigh-Taylor instability, increasing the amplitude and scale of the
first instabilities to form.  This effect is shown for s15A in Figure
\ref{s15A_pert}.  The case of $0.5\%$ random perturbations to the velocity
results in a scale for the initial Rayleigh-Taylor instabilities that
is roughly equivalent to the case where the only perturbations were
those arising from the Cartesian grid itself.  Other models had similar
resolution, such that grid perturbations were roughly equivalent to
velocity perturbations of $0.5\%$ and velocity perturbation of $2.0\%$
had a noticeable effect on the initial scale of the instability.
Perturbations arising from the grid were no larger than $2\%$, well
within the regime of perturbations expected to arise from convection.

Because the zD-series models were more compact than sA-series models,
the reverse shock reached the centers of zD models faster than in the
sA models.  This had the effect of shutting off mixing in the
zD-series before the Rayleigh-Taylor instability had time to become
fully non-linear \citep[see also][]{Herant&Woosley:1994}.  The initial
perturbations had a greater effect on the final state of the zD
simulations, while the initial perturbation scale and spectrum are
washed out in the sA models as a result of their longer mixing times.
A simulation with random velocity perturbations of $5\%$ was performed
for Model z25D.

Models were initially mapped onto the two-dimensional grid such that
their inner iron cores were resolved with at least 4 blocks of 16
zones each.  This was sufficient to ensure that the rest of the star
was accurately resolved.  As the simulation progressed and the model
stars expanded, the maximum refinement level of the simulation was
turned down, so that the model star was always resolved at about the
same percentage of the radius of its inner core of \He \  and heavier
isotopes.  This is the region where Rayleigh-Taylor mixing takes
place. The star expands homologously, ensuring that all regions will
be adequately resolved.

The solar composition models were mapped onto a grid $5\times 10^{14}$
cm on a side.  The zero-metallicity stars were mapped onto a grid
$1.4\times 10^{14}$ cm on a side.  The portion of the grid outside the
original \code{KEPLER} model was initialized with a density
proportional to $r^{-2}$.  Simulations with an outer density
proportional to $r^{-3.1}$ was also performed for Models s15A, s25A,
and z25D.  These density profiles span the realistic range of smooth
density distributions outside real stars.  No difference in the final
profiles for density, temperature, pressure, or composition was found
between models with different outer density profiles. The density
profile of the surrounding material therefor has no effect on the
amount of Rayleigh-Taylor mixing that goes on inside the star,
provided very little mass as a proportion of the original mass of the
star is swept up in the first days of the explosion.  The amount of
mass added to the grid from ambient density was $< 2\%$  for all models.

Calculations were run at least until the Rayleigh-Taylor fingers had
ceased to move with respect to the mass coordinate of the star. This
happened ~2 hours after core bounce for Model z15D, ~4 hours after
core bounce for z25D, and ~ 7 days after core bounce for the s15A and
s25A models.  All models were followed to $10^6$ seconds, long after
Rayleigh-Taylor mixing had frozen out in the zero metallicity stars,
but long enough that infall though the inner boundary had reached an
asymptotic stage and the final remnant mass from these two-dimensional
simulations could be determined.

\section{RESULTS}
\lSect{results}

\begin{figure}
  \includegraphics[width=0.45\textwidth]{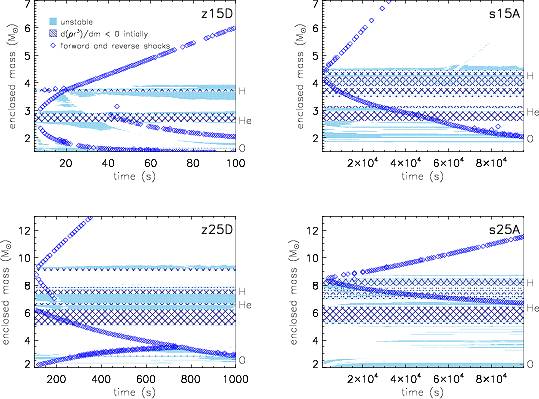}
  \caption{Stability evolution for the first $\approx5\%$ of
    simulation time.  Left axes show the location of the base of the
    shell indicated.  Note different time and mass axes scale between
    the models.  The reverse shock takes far longer to propagate back
    for solar composition than for Pop III stars. A greater portion of
    mass in the solar stars is RT unstable, leading to more mixing in
    these stars.\label{stability}}
\end{figure}

Figure \ref{stability} shows the evolution of the stable and unstable
regions of the models, and the position in mass coordinate of the
forward and reverse shocks.  A reverse shock forms when the outgoing
shock encounters a region of increasing $\rho r^3$
\citep{Herant&Woosley:1994,Woosley&Weaver:1995}.  When the shock encounters a density profile that
falls off with a flatter slope than $\propto r^{-3}$, i.e. when it
encounters a region of increasing $\rho r^3$ it decelerates.  The
deceleration of the forward shock front reverses the direction of the
pressure gradient, which slows down the layers interior to the shock,
as well.  Shocked material piles up and forms a high density
post-shock shell. The reverse shock forms at the inner boundary of the
high-density shell of decelerated matter and propagates down into the
star, toward its center, slowing down the deeper, inner layers of the
star \citep{Kifonidis:2003}. The deceleration of the shock creates a steep
pressure gradient in the opposite direction to the existing
gravitational and density gradients.  This pressure gradient can
overwhelm the gravitational gradient, and in doing so triggers the
formation of Rayleigh-Taylor instabilities in the material.  The
Rayleigh-Taylor instability develops until the reverse shock has
passed by, at which point the material becomes stable again, and the
instabilities cease to grow exponentially.  Figure \ref{stability}
covers the period of time from when the models were first mapped to
\code{FLASH} to slightly beyond the time when the Rayleigh-Taylor
instability fingers began to grow.

\begin{figure}
  \includegraphics[width=0.45\textwidth]{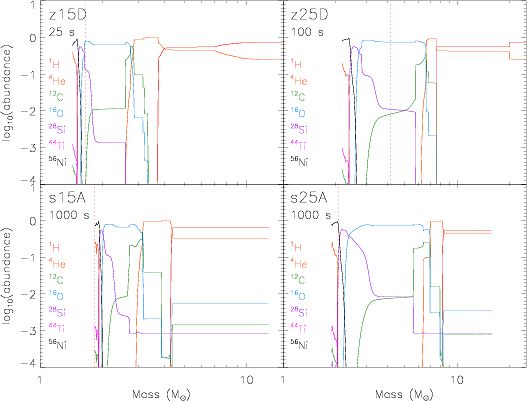}
  \caption{Distribution of isotopes just after core bounce as a
    function of mass coordinate.  The dotted vertical line shows the
    position of the mass cut--matter to the left of this line will fall
    back onto the star.  Without mixing, no \Ni \ escapes from stars of
    primordial composition, while most of the $^{56}$Ni core is
    expelled from the solar composition
    supernovae. \label{compare_el_prof_orig}}
\end{figure}

\begin{figure}
  \includegraphics[width=0.45\textwidth]{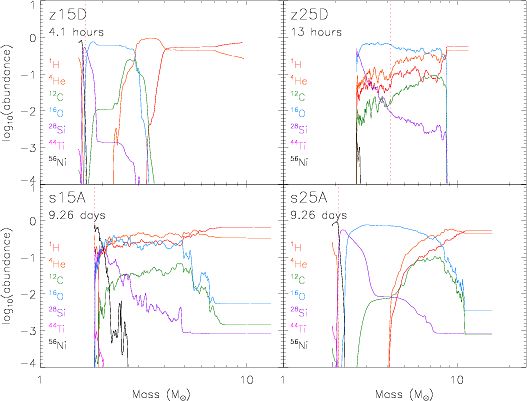}
  \caption{Distribution of isotopes after mixing has ceased.  Note that
    we find almost no difference between this distribution and the
    initial configuration for Model z15D.  Mixing is confined to the
    O-He shells for Model z25D.  Mixing has penetrated the Si/O layer
    for Model s15A. \label{compare_el_prof_flash}}
\end{figure}

\begin{figure}
  \includegraphics[width=0.45\textwidth]{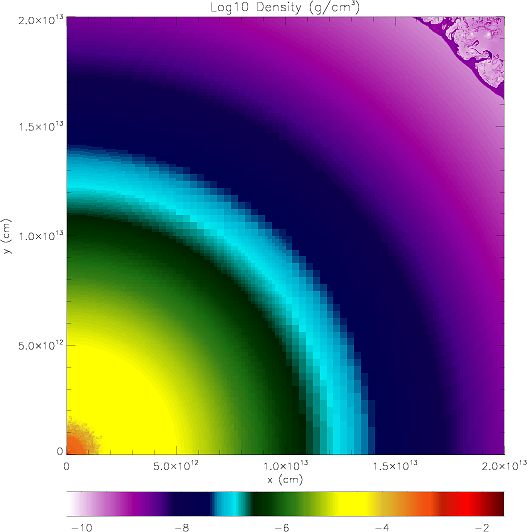}
  \caption{Log(density) snapshot of Model z15D after mixing has
    ceased.  The RT instability visible in the upper right hand corner
    is most likely a result of the artificial outer density profile.
    It represents the outer layer of the star.  The mixed inner layer
    extends only about $1/10$ of the radius of the star, covering a
    much smaller region than in the solar models.\label{z15D_dens}}
\end{figure}

\begin{figure}
  \includegraphics[width=0.45\textwidth]{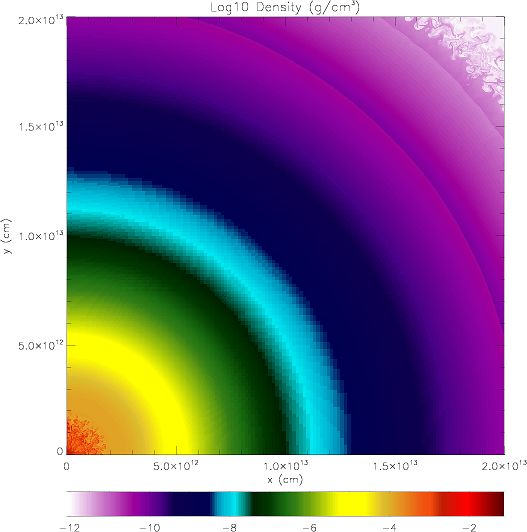}
  \caption{Log(density) snapshot of Model z25D after mixing has
    ceased. Like Model z15D, a small portion of the star is mixed when
    compared to the corresponding model of solar composition.  The RT
    instabilities visible in the upper corner formed at the boundary
    of the star with an artificial density background.
    \label{z25D_dens}}
\end{figure}

\begin{figure}
  \centering
  \includegraphics[width=0.225\textwidth]{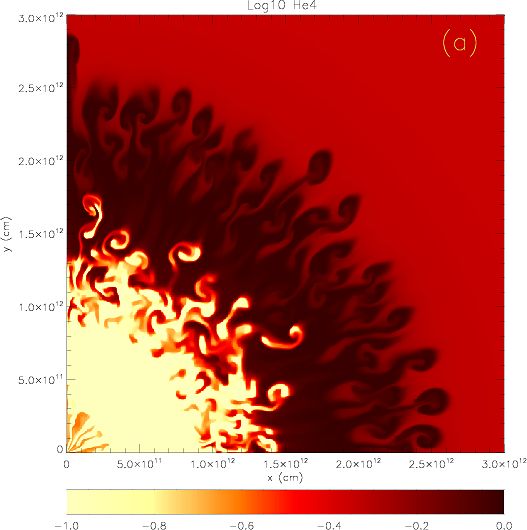}
  \includegraphics[width=0.225\textwidth]{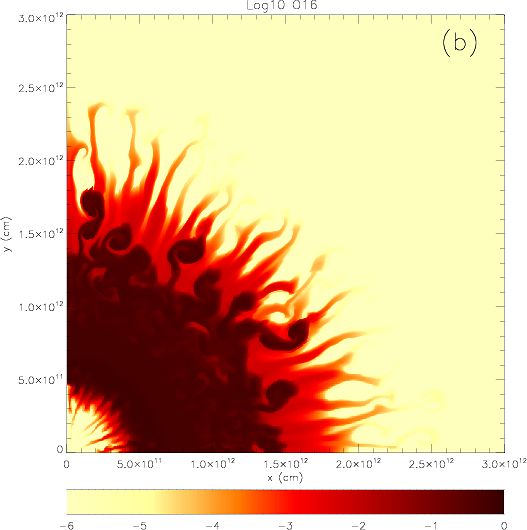}
  \includegraphics[width=0.225\textwidth]{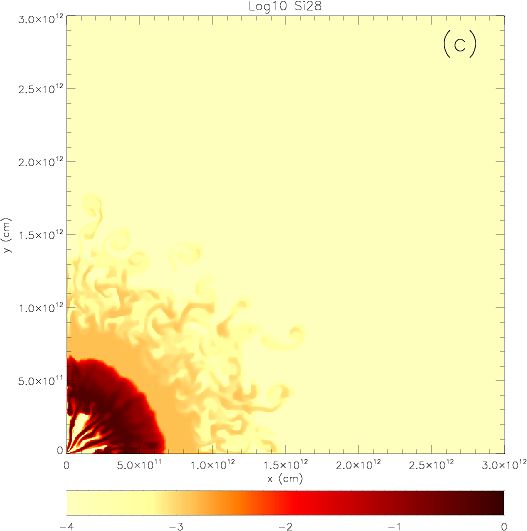}
  \includegraphics[width=0.225\textwidth]{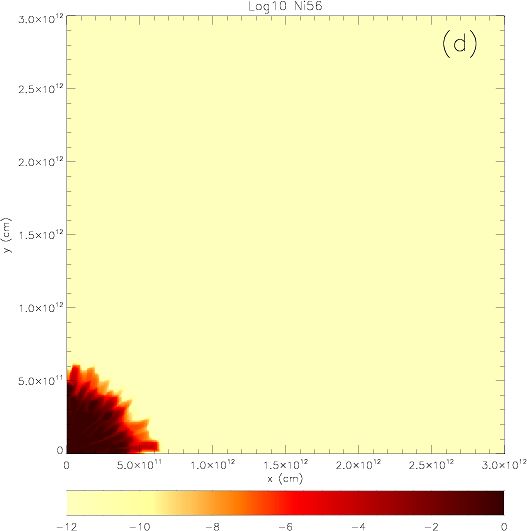}
  \caption{Final distribution of isotopes in the inner core of Model
    z15D.  Note that the mixed region extends only to about 1/10 of
    the radius of the star--a much smaller proportion than in the
    solar metallicity stars. Pictured is the run with a $2\%$
    percent perturbation in velocity.  The Rayleigh-Taylor
    instabilities have had a shorter time to grow, and so have
    retained more of their original shape, than in the solar
    metallicity models.\label{z15D_all}}
\end{figure}

\begin{figure}
  \centering
  \includegraphics[width=0.225\textwidth]{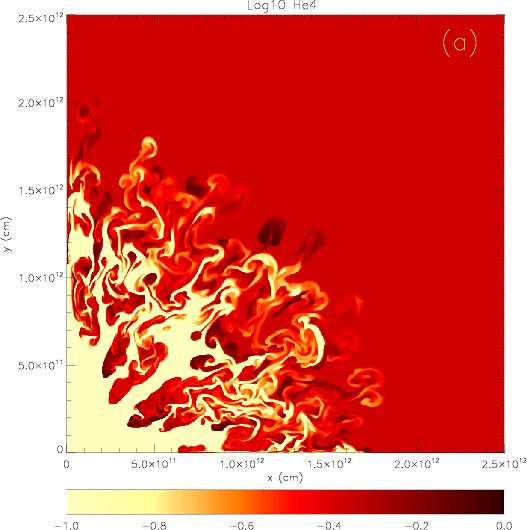}
  \includegraphics[width=0.225\textwidth]{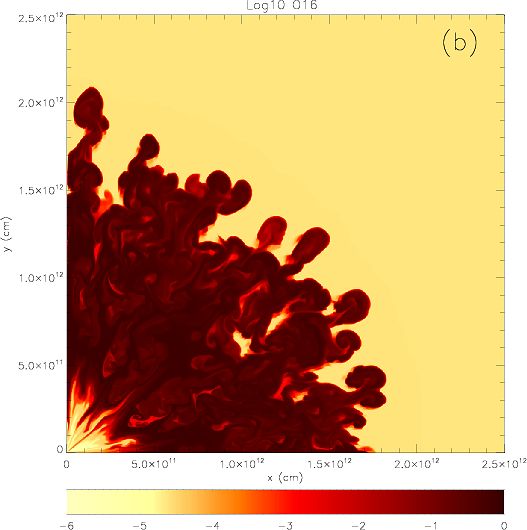}
  \includegraphics[width=0.225\textwidth]{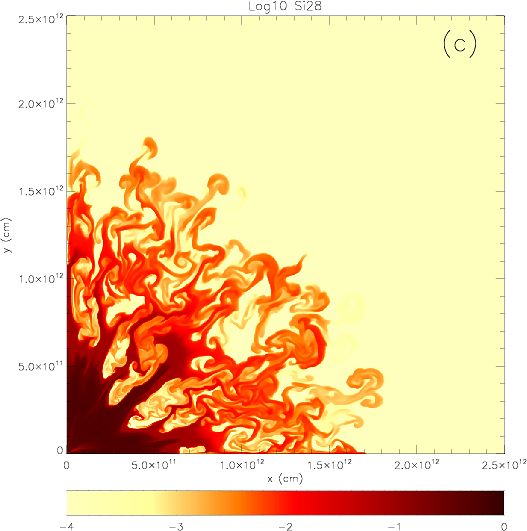}
  \includegraphics[width=0.225\textwidth]{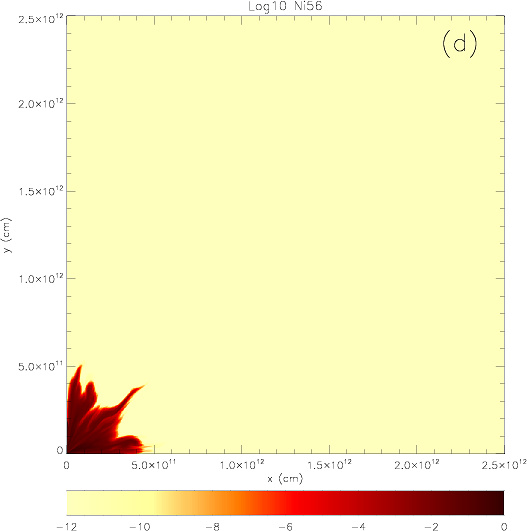}
  \caption{Final distribution of isotopes in the inner core of model
    z25D with a $5\%$ perturbation in velocity. Note the mixed
    region goes out to only about 1/10 of the radius of the star--a
    much smaller proportion than in the solar metallicity stars,
    although the mixed region is greater and the initial scale of
    the perturbation is less visible than in Model z15D. The reverse
    shock took longer to pass though Model z25D then Model
    z15D.\label{z25D_all}}
\end{figure}

Figure \ref{stability} shows that the Rayleigh-Taylor instability had
far more time to develop in the solar composition models than
primordial composition models.  The zero-metallicity models,
particularly Model z15D, were far more compact than their solar
metallicity counterparts.  The reverse shock took only $\approx$100
seconds to reach the center of Model z15D, leaving very little time
for the Rayleigh-Taylor instability to grow.  This is reflected in the
low degree of mixing seen in Figure \ref{compare_el_prof_flash}, which
shows the final distribution of isotopes as a function of mass for all
stars.  The original structure of these stars can be seen in Figure
\ref{compare_el_prof_orig}.  Two-dimensional snapshots of mixing are
shown in Figure \ref{z15D_dens}, which shows the density structure of
the entire star, and Figure \ref{z15D_all}, which shows the isotopic
composition of the mixed region at the center of the model.  In Model
z25D, the reverse shock took longer, about 10$^3$ seconds, to reach
the center of the star, allowing the RT instability to grow for a
longer period of time.  Figure \ref{stability} shows an unstable band
between the $^4$He/$^{12}$C-$^{16}$O shell boundary, and that was
indeed the place we saw mixing in these models.  More mixing in z25D
than z15D can be seen in Figures \ref{compare_el_prof_flash},
\ref{z25D_dens}, and \ref{z25D_all}.

\begin{figure}
  \includegraphics[width=0.45\textwidth]{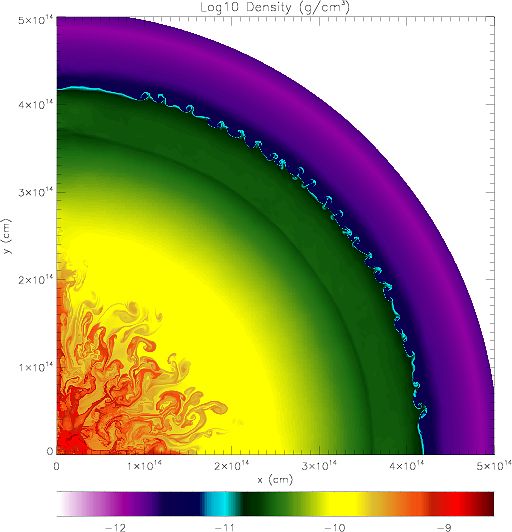}
  \caption{Log(density) snapshot of Model z25D after mixing has
    ceased. The Rayleigh-Taylor instability that began at the O/He
    interface has mixed the inner layers of the star down to the iron
    core. The flow visible along the y-axis is a result of dimensional
    splitting in the hydrodynamic solver.\label{s15A_dens}}
\end{figure}

\begin{figure}
  \includegraphics[width=0.45\textwidth]{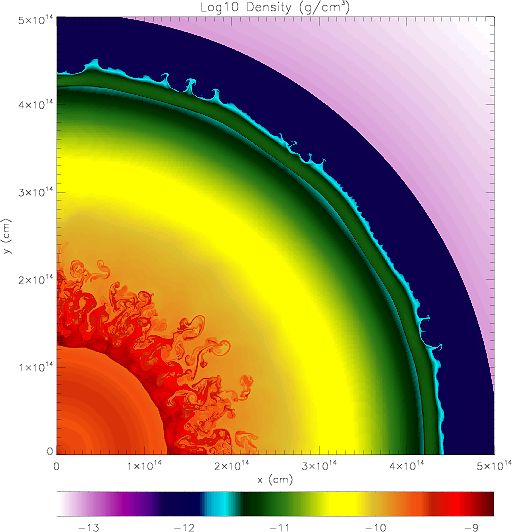}
  \caption{Log(density) snapshot of Model s25A after mixing has
    essentially ceased. The Rayleigh-Taylor instability that began at
    the O/He interface has not penetrated through the thick $^{16}$O[
      shell, as it had with model s15, where the shell was thinner (Figure
      \ref{s15A_dens}.) \label{s25A_dens}}
\end{figure}

Solar composition models were about 50 times larger in radius than
their primordial composition counterparts, and correspondingly less
dense.  The reverse shock took longer to form and 10$^5$ and
$2\times10^5$ seconds to propagate back to the mass-coordinate origin
for Models s15A and s25A, respectively.  This was about 100 times
longer than for Model z25D, giving the Rayleigh-Taylor instability
more time to develop.  Additionally, a wider range of regions between
and including the \He \ and Si+S shells were unstable than in the zero
metallicity counterparts. Figure \ref{stability} implies that the
solar metallicity stars were mixed to a greater degree than stars of
zero metal initial composition.  Figures \ref{s15A_dens} and
\ref{s25A_dens} show that about half the solar composition stars were
mixed, compared to only about 1/10 of the zero metallicity stars.

\subsection{Infall}
\lSect{infall}

\begin{figure}
   \includegraphics[width=0.45\textwidth]{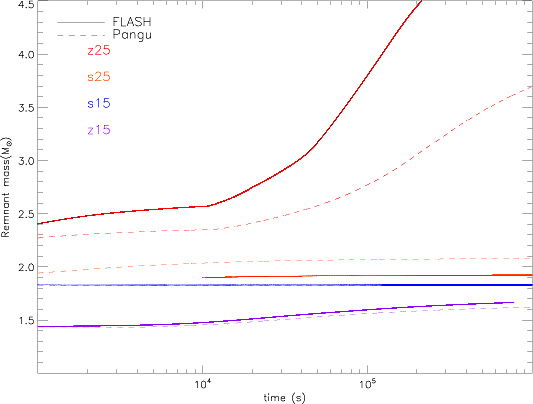}
   \caption{Comparison of infall rates through the inner boundary for
     1-d \code{PANGU} and 2-d \code{FLASH} Eulerian codes.  For all
     stars except Model s15A, \code{FLASH} overestimates the initial
     infall rate, which is reflected in a larger remnant mass.  This
     is due to limited resolution at the origin in the \code{FLASH}
     code.  The shape of the \code{FLASH} infall curves matches the
     shape of those curves in \code{PANGU}.  The exception is Model
     z25D, which shows a dramatic departure from the shape of the
     \code{PANGU} curve at later times.  This is probably because the
     extra remnant mass is large enough to effect the infall rate
     through the inner boundary.  \label{infall}}
 \end{figure}

Infall is as important as mixing for determining the final yield of a
supernova.  For a freely expanding supernova remnant, the rate at
which mass accretes onto the black hole or neutron star at the center
of the explosion is given by $\dot{M} \propto t^{-5/3}$ \citep{Chevalier:1989}.
It takes supernova explosions on the order of $10^6$ seconds to reach
this asymptotic, freely expanding stage, at which time it is possible
to determine the final mass of the stellar remnant by extrapolating
from the asymptotic infall rate.  Our final remnant masses are
compared with those found with one-dimensional calculations carried
out by \citet{Zhang:2008} with the \code{PANGU} code in Figure
\ref{infall}.

\code{PANGU} is a one-dimensional hydrodynamics code based on the
second-order semi-discrete finite difference central scheme of
\citet{Kurganov:2000}.  Time evolution was carried out by a third-order total
variation diminishing Runge-Kutta method \citep{Shu:1989}.  \citet{Zhang:2008}
simulated the explosions of stars in the \citet{Woosley&Heger:2007} and
\citet{Heger&Woosley:2008} surveys.  Using this one-dimensional code, the authors
followed the evolution of the supernova remnant out to $~10^6$
seconds, to the time at which the accretion rate onto the central
remnant had reached an asymptotic dependence on time and the final
remnant mass could be determined.

\begin{figure}
  \centering
  \includegraphics[width=0.225\textwidth]{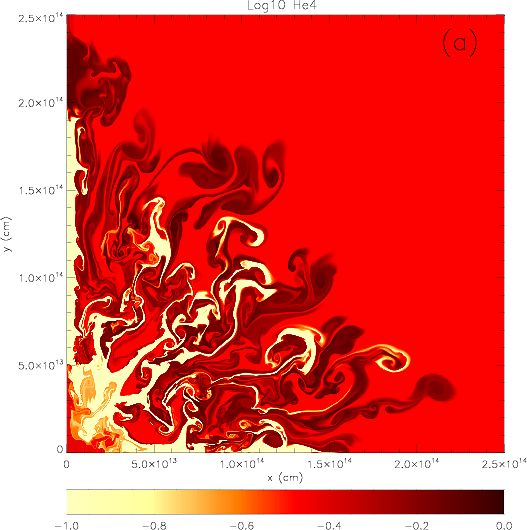}
  \includegraphics[width=0.225\textwidth]{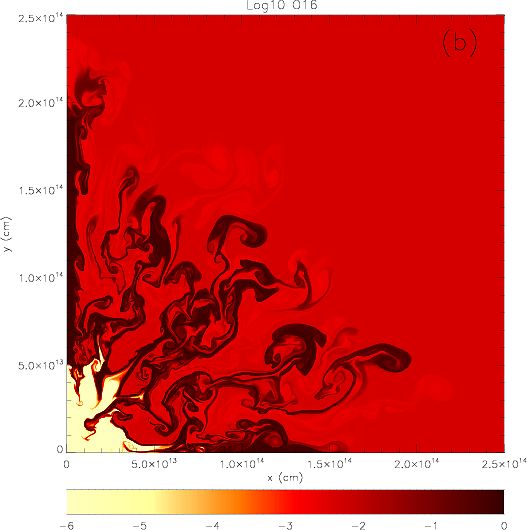}
  \includegraphics[width=0.225\textwidth]{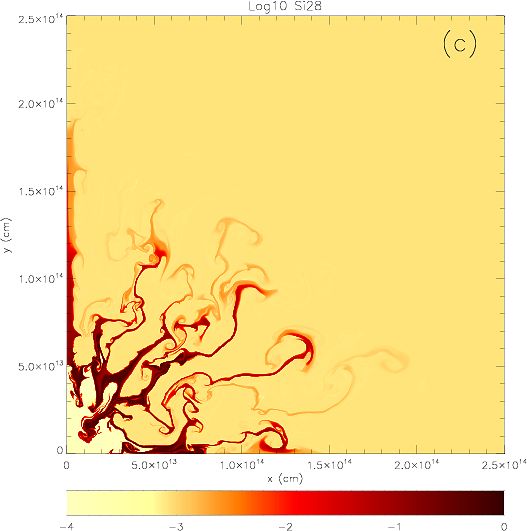}
  \includegraphics[width=0.225\textwidth]{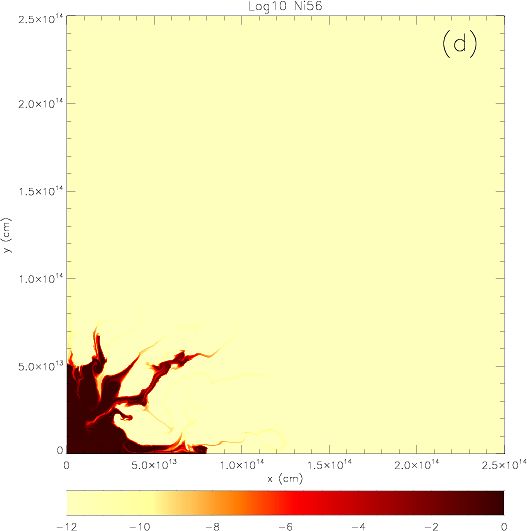}
  \caption{Final distribution of isotopes in the inner core of Model
    s15A.  The helium, oxygen, and silicon shells have been
    disrupted by the Rayleigh-Taylor instability, which has mixed
    $^{56}$Ni out of the core and $^1$H all the way in to the inner
    layers.  Model s15A is more completely mixed than any other
    model presented in this paper, because of both the long
    timescale for the reverse shock and a thinner oxygen layer.
    \label{s15A_all}}
\end{figure}

\begin{figure}
  \centering
  \includegraphics[width=0.225\textwidth]{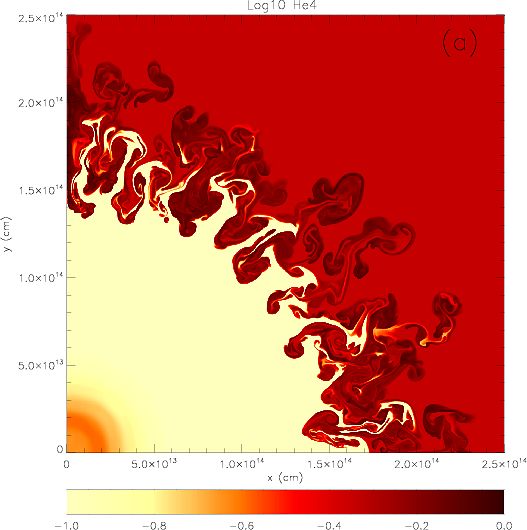}
  \includegraphics[width=0.225\textwidth]{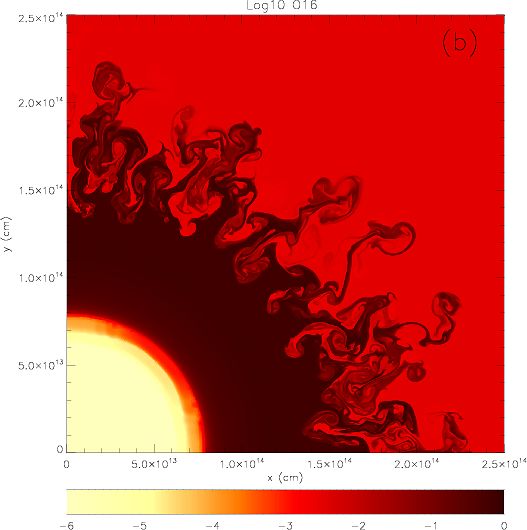}
  \includegraphics[width=0.225\textwidth]{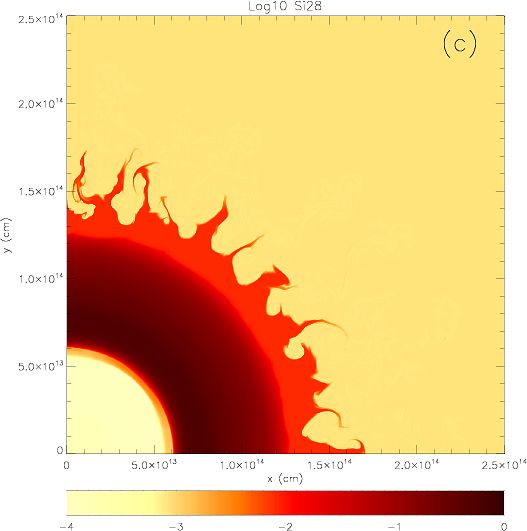}
  \includegraphics[width=0.225\textwidth]{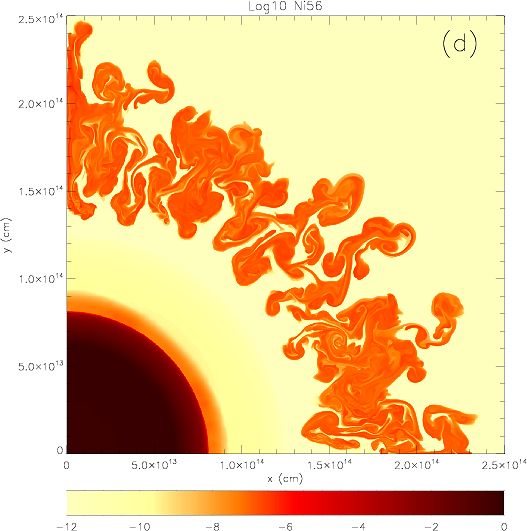}
  \caption{Final distribution of isotopes in the inner portion of Model
    s25A. The $^4$He layer has been completely disrupted, though the
    RT instability has not penetrated past the $^{16}$O layer.  The
    ring of $^{56}$Ni visible was formed by explosive nuclear
    burning. $^{56}$Ni has not been mixed out of the center of the
    star, and the $^{28}$Si layer is only marginally effected by the
    Rayleigh-Taylor instability.  \label{s25A_all}}
\end{figure}

Our final remnant masses showed good agreement with the \code{PANGU}
results for all Models except Model z25D, as shown in Figure\ref{infall}.
The shape of the infall curves for our stars, as shown in Figure
\ref{infall}, match the \code{PANGU} results well, though the remnant
masses from \code{FLASH} are larger except in the case of s25A.  Model s25A
was mapped to \code{FLASH} after $1 \times 10^4$ seconds of evolution
in \code{KEPLER}.  Although this does not alter the final mixed state of
the star, it does alter the amount of mass that accumulates at the
inner boundary, since most of the infall occurs during the first
$10^4$ seconds, when the star was still being evolved forward with the
Lagrangian code \code{KEPLER}.  Both 15 solar mass models are in good
agreement with the \code{PANGU} results.  z25D shows the most
deviation, most likely because this model experienced the most fall
back of any of the models studied here.  We were unable to replicate
the fine resolution employed by a one dimensional code, and that is
almost certainly the reason for the enhanced infall mass.  The initial
remnant mass in 2D was sufficiently larger than the 1D remnant mass that
the additional force it exerted on the surrounding material was large
enough to cause additional infall.  This led to a larger remnant,
causing a mild runaway effect.  While the remnant mass obtained with
\code{FLASH} for Model z25D is likely inaccurate, Rayleigh-Taylor
mixing in this star has stopped by $2 \times 10^4$ seconds, at which
point the remnant mass in the \code{FLASH} simulation was not
significantly larger than that in the \code{PANGU} simulation. It is
unlikely that increased infall has had a significant impact on the
evolution of the Rayleigh-Taylor instability, so the mixing results
remain sound.

\subsection{Perturbations}
\lSect{perturbations}

\begin{figure}
  \includegraphics[width=0.45\textwidth]{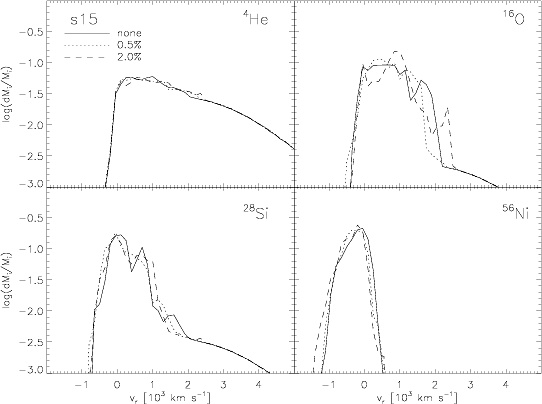}
  \caption{Mass fraction of chemical isotopes as a function of radial
    velocity for model s15A.  Differences between the models are
    larger than for Model s25A (see Figure \ref{s25_changevel}, but
    are not systematic with perturbation, and become significant only
    below mass fractions of around $5\%$ of the mass of \Ox and
    \Si.  \label{s15_changevel}}
\end{figure}

\begin{figure}
  \includegraphics[width=0.45\textwidth]{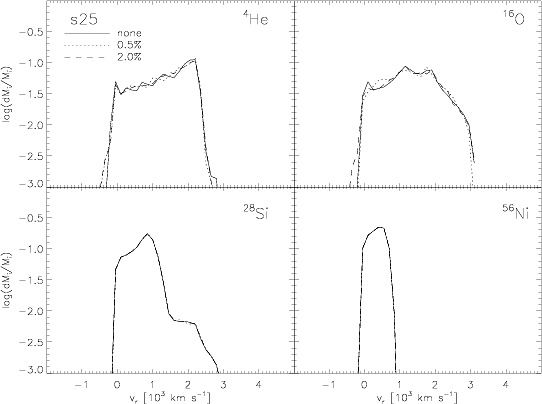}
  \caption{Mass fraction of chemical isotopes as a function of
    radial velocity for Model s25A with different perturbations.
    There are only small differences, and the differences are not
    systematic with the amount of perturbation imposed on the initial
    model.\label{s25_changevel}}
\end{figure}

\begin{figure}
  \includegraphics[width=0.45\textwidth]{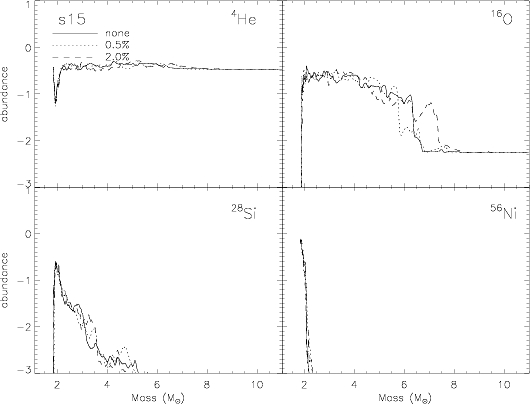}
  \caption{Isotopic distribution as a function of mass for Model s15A
    with different initial perturbations.  The isotopic distribution
    is not systematic with the amount of perturbation.  \Ox shows the
    greatest change, but only at mass fractions below
    $10\%$.  \label{s15_changeel}}
\end{figure}

\begin{figure}
  \includegraphics[width=0.45\textwidth]{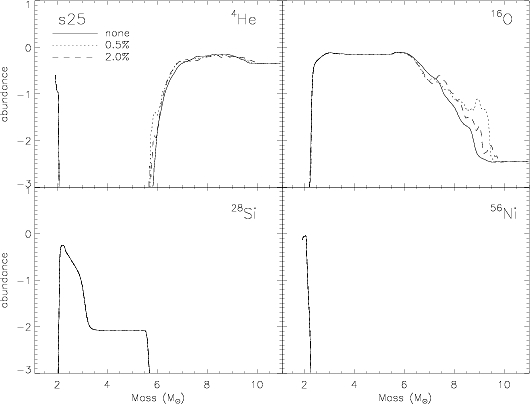}
  \caption{Isotope distribution as a function of mass for Model s25A
    with different initial perturbations.  Where the distributions of
    isotopes are different, they are not systematic with perturbation.
    Differences between the runs are also small, and only become
    obvious for \Ox, with oxygen concentrations are less than
    $10\%$. \label{s25_changeel}}
\end{figure}

Random perturbations of amplitude $0.5\%$ and $2\%$ of the original
velocity profile were added to the initial model for the solar
composition models, as described in \Sect{FLASH}.  Random
perturbations of $2\%$ were added to the zero-metal stars, and an
additional simulation with an initial perturbation of $5\%$ was
performed for Model z25D. Perturbations of $0.5\%$ produce an original
size scale for the Rayleigh-Taylor instabilities that is about the
same size as those arising for the non-perturbed case, while the $2\%$
perturbations result in a larger initial scale for the instability,
implying that grid perturbations for these models are between $0.5\%$
and $2\%$, as shown in Figure \ref{s15A_pert}.  The Rayleigh-Taylor
instabilities can grow for at least as long as the reverse shock takes
to reach the origin.  The instabilities in the solar metallicity stars
can grow for many $e-$folding times, a long enough time to wash out
the initial scale and spectrum of the instabilities.  The final states
of the perturbed models appear essentially identical to the
unperturbed models.  The distribution of the isotopes in both velocity
(Figures \ref{s15_changevel} and \ref{s25_changevel}) and and mass
space (Figures \ref{s15_changeel} and \ref{s25_changeel}) has no
systematic correlation with the magnitude of the initial perturbation
for solar metallicity stars.  The scale of the perturbations one would
expect in a real star is set by the magnitude of the convective
velocities, which are on the order of $0.5\%$ of the total velocity.
\begin{figure}
  \includegraphics[width=0.45\textwidth]{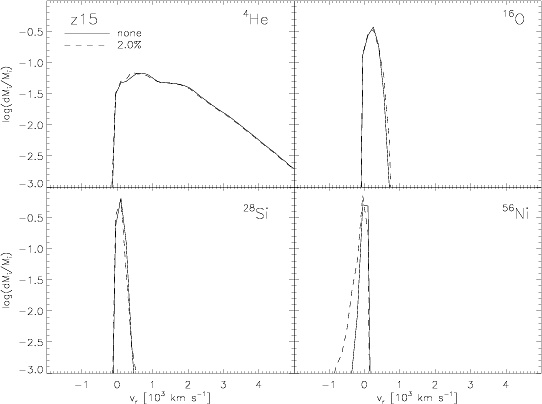}
  \caption{Mass fraction of chemical isotopes as a function of radial
    velocity for Model z15D for different perturbations.  Larger
    initial perturbations lead to wider distributions. These
    differences effect about $10\%$ of a given
    isotope.\label{z15_changevel}}
\end{figure}

\begin{figure}
  \includegraphics[width=0.45\textwidth]{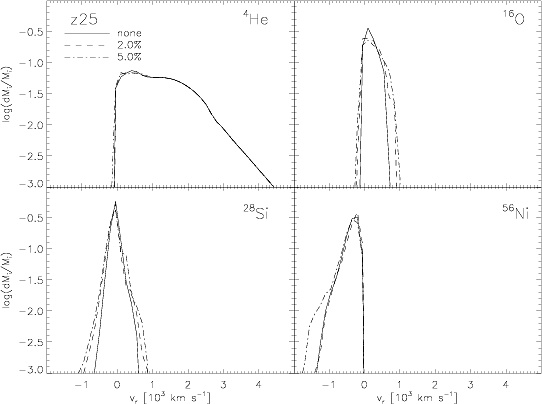}
  \caption{The same as Figure \ref{z15_changevel}, but for model z25D.
    As with Model z15D, a greater perturbation results in slightly
    enhances negative velocities, especially for \Ni.
    \label{z25_changevel}}
\end{figure}

\begin{figure}
  \includegraphics[width=0.45\textwidth]{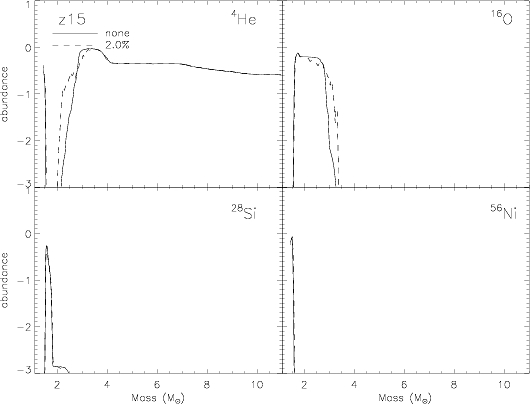}
  \caption{Isotopic distribution as a function of mass for Model z15D
    with different initial perturbations.  A larger perturbation leads
    to more mixing. \label{z15_changeel}}
\end{figure}

\begin{figure}
  \includegraphics[width=0.45\textwidth]{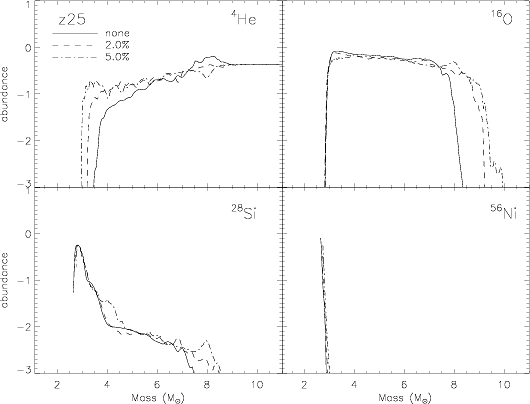}
  \caption{Isotopic distribution as a function of mass for Model z25D
    with different initial perturbations.  The distribution changes
    with the level of perturbation--larger perturbations lead to more
    mixing in a systematic way, and the difference between different
    perturbations is visible at the level of $50\%$ mass fraction of
    \He \ and \Ox, much larger than in the more non-linear solar
    metallicity stars.\label{z25_changeel}}
\end{figure}

Perturbations for the zero-metallicity case have a greater effect on
the final amount of mixing in these stars.  Larger perturbations lead
to a larger size scale for the initial Rayleigh-Taylor fingers, which
allows them to grow more quickly before the reverse shock passes them
and the pressure gradient is no longer opposite the density gradient.
In these models, the Rayleigh-Taylor instabilities cannot grow for
many $e-$folding times, and so the initial scale of the perturbations
matters.  This can be seen in Figures \ref{z15_changeel} and
\ref{z25_changeel}, which show the final distribution of isotopes as a
function of enclosed mass for different amounts of perturbation.  The
final amount of mixing is set by the size scale of the initial
perturbation.  The amount of mixing determines the distribution of
isotopes with velocity, as well, as shown in Figures
\ref{z15_changevel} and \ref{z25_changevel}. \Ox \ and \He \ are the
most affected. The distance in mass space over which these isotopes
are mixed varied by around 1 \Msun \ for the perturbed models.

\subsection{Velocity Distribution}
\lSect{velocity}

\begin{figure}
  \includegraphics[width=0.45\textwidth]{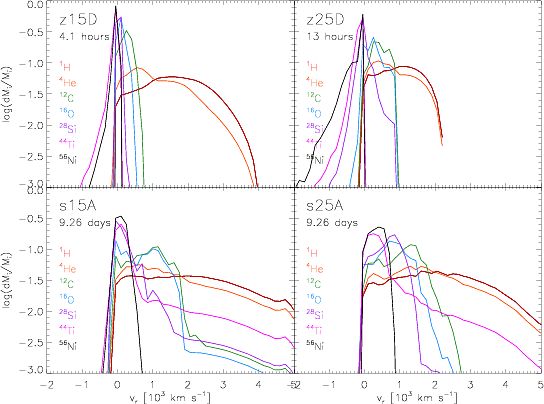}
  \caption{Mass fraction of chemical isotopes as a function of
    velocity for Models s15A and s25A at $t=9.26$ days and z15 and z25
    at $t=2$ and 4 hours, respectively.  Negative velocities indicate
    material falling toward the point mass at the center of the
    star. Virtually no iron-peak elements escape from the primordial
    composition stars, and their velocity distributions are smaller
    than the solar models.  The peaks in the velocity distribution for
    $^{56}$Ni for the solar stars are not high enough to match what
    was observed in 1987A, but are consistent with previous studies of
    mixing in RSGs.\label{velvsel}}
\end{figure}

The velocity distribution of chemical isotopes is shown in Figure
\ref{velvsel}.  $^{44}$Ti was mixed to relatively high velocities in
our solar metallicity stars. The solar stars show \Hy \ and \He \  at high
velocities of $4\times10^3$ km s$^{-1}$.  These isotopes are also
mixed all the way to the core of the solar metallicity stars, and some
fraction of them reaches negative velocities of less than $-0.5 \times
10^3$ km s$^{-1}$. In Model s15A, where the most mixing takes place,
we see $^{16}$O and $^{12}$C mixed out to $2 \times 10^3$ km s$^{-1}$
in velocity space.  \Ni \  does not reach the high velocities observed in
1987A.  Model s15A shows a peak in the \Ni \  velocity at $\approx 0$ km
s$^{-1}$, with a tail extending to $0.7 \times 10^3$ km
s$^{-1}$. Model s25A shows slightly higher velocities, with the \Ni
distribution peaking at around $0.4 \times 10^3$ km s$^{-1}$ and
reaching out to $0.9\times 10^3$ km s$^{-1}$.  These higher velocities
for \Ni \  are probably due to smaller amounts of fallback in this star.

The zero metallicity stars show lower velocities overall.  At the time
they exploded as supernovae, these stars were more compact than solar
composition stars.  The supernova shock ran into a higher fraction of
the total mass of the star at an earlier time.  In neither zero metal
model were the heavier isotopes mixed out to higher velocities than $1
\times 10^3$ km s$^{-1}$, but in Model z15D the velocities of \Ni \  and
$^{44}$Ti had slight positive components, whereas in Model z25D the
velocity distribution of these isotopes was almost entirely negative.
No $^{56}$Ni or $^{44}$Ti escape from Model z25D.

\citet{Kifonidis:2006} in their 1987A-type models, which had a metallicity and
mass intermediate to the stars studied in this paper, saw a well-mixed
heavy element core. The velocity profiles for isotopes from $^{16}$O
to the iron group were very similar.  Because we did not see the iron
core of our solar metallicity stars mixed to the degree seen in
\citet{Kifonidis:2006}, our $^{56}$Ni had a slower velocity distribution, while
lighter isotopes, from \Ti \ and \Si \ and lighter, were skewed
towards higher velocities.

The \Ni \  velocities observed in the zero-metallicity models were lower
than those produced in earlier attempts at modeling SN 1987A, as well.
The maximum iron-peak velocities obtained was on the order of 1300 km
s$^{-1}$ \citep{Arnett:1989,Hachisu:1990,Fryxell:1991,Muller:1991, Herant:1991}. In particular,
\citet{Herant:1991} found that following the radioactive decay of \Ni \  and
\Co \ increased the velocity of iron-peak elements slightly, but not
enough to match observations of SN 1987A.  Our iron-peak velocities
were lower still.  This was partly due to the large amount of fallback
experienced by the primordial composition models.  Most of the \Ni
fell through the inner boundary of our zD-series simulations before it
had time to power a nickel bubble like those seen in SPH simulations
of SN 1987A \citep{Herant:1991}.  Our progenitors were also more compact,
with smaller helium cores, than the progenitor models for 1987A used
in the above simulations.

$^{56}$Ni was not mixed out to the the high velocities seen in
SN1987A.  \citet{Utrobin:2004} report that $^{56}$Ni was mixed out to at least
$2.5 \times 10^3$ km s$^{-1}$.  In no model did we see nickel at these
high velocities.  Significant low-order asymmetry in the explosion is
probably necessary to reach \Ni \  velocities high enough to match
observations.

In their 1994 paper on mixing in supernovae with red supergiant
progenitors, which employed progenitor models similar to the s15A and
s25A progenitor models in this survey, Herant and Woosley found a
similar distribution of isotopes in velocity space to the results
presented here.  They found a peak in the velocity distribution of
$^{16}$O at around $1.4 \times 10^{3}$km s$^{-1}$ for their 25 \Msun
\ star, we see one at around $1.1\times10^{3}$ km s$^{-1}$.  We see
oxygen mixed out to slightly higher velocities in our model s15A than
are given in \citet{Herant&Woosley:1994}.  Their models show no negative velocities,
which is probably a result of the SPH technique employed, which did
not allow for accretion onto a sink particle at the center of the
simulation.

\subsection{Yields}
\lSect{yields}

The yields of our stars are greatly affected by the amount of fallback
and degree of mixing induced by Rayleigh-Taylor instabilities.  For
models that experienced a great deal of fallback, as in case of the
zero-metallicity models, mixing is of crucial importance in
determining the final yields.  Figure \ref{compare_el_prof_orig} shows
the original distribution of isotopes in the four stars studied in
this paper.  The distribution of isotopes after all mixing has ceased
is shown by Figure \ref{compare_el_prof_flash}.  As Figure
\ref{compare_el_prof_flash} and Figures \ref{z15D_all} and
\ref{z25D_all} show, layers interior to oxygen in the zero-metallicity
stars have not mixed appreciably.  A far larger portion of the solar
metalicity stars has been mixed together.  The Rayleigh-Taylor
instability has penetrated as far as the \Ni \ core of model s15A,
completely disrupting the exterior shells.  This can be seen in Figure
\ref{s15A_all}.  Figure \ref{s25A_all} also shows that Model s25A
experienced more mixing than Model z25D (Figure \ref{z25D_all}), but
not quite as much as Model s15A.  The lack of mixing into the interior
layers of the zero-metallicity models means that very little of the
silicon shell, and virtually none of the elements interior to this
shell are mixed out.  This, coupled with a large amount of infall,
means that these Pop III stars expel almost no $^{56}$Ni.

To eliminate the possibility that the absence of Rayleigh-Taylor
induced mixing at the edge of the $^{56}$Ni shell in Models z15D and
z25D could have been due to mapping to two dimensions at too late a
time we performed four simulations of the interior of Model z25D with
an inner boundary of $1 \times 10^9$ cm, an order of magnitude smaller
than that used to simulate the star in the main simulations presented
in this paper.  These simulations were run at ten times the resolution
of the simulations presented in \Sect{calculations}.  \code{KEPLER}
models were mapped in 10 and 40 seconds after bounce.  To maintain
consistency with the simulations that used a mapping from \code{KEPLER}
at 100 s after bounce, we initially used a reflecting condition at the
inner boundary, which did not allow infall.  At 100 seconds after
bounce for all of these high resolution models the inner boundary
condition was changed to be zero-gradient.  No Rayleigh-Taylor
instabilities are seen to develop by 1$\times 10^3$ s of simulation
time, the point when the instabilities are clearly defined and
growing along the C-O/He interface.  Perhaps the perturbations from
the Cartesian grid at this resolution are too small to effectively
seed the instability.  We perturbed the velocity on the grid with an
$l=8$ order Legendre polynomial with an amplitude $10\%$ of the
initial velocity.  While this is probably unphysical, it represents a
limiting case: The absence of Rayleigh-Taylor induced mixing with a
perturbation this extreme implies its absence in the larger, lower
resolution simulations is reasonable. The first $1 \times 10^3$ s of
the simulation were run at this high resolution.  By this point the
Rayleigh-Taylor instabilities are clearly visible at the C-O/He
boundary, but there is no hint of the development of this instability
at the interface between the $^{56}$Ni core and the overlying layers.
\citet{Kifonidis:2006} speculated that the reason they saw mixing at the Si/O
boundary when earlier studies did not was the ''ad hoc''
initialization of these explosions with a piston or thermal bomb,
insufficient resolution, or differences in structure between their
progenitor models and those employed in previous studies.
Insufficient resolution does not appear to be the case here, and our
extreme perturbations imply that seeding perturbations through
neutrino-driven convection by starting the simulation earlier would
have little effect.  We can only conclude that this is a valid result
arising from the structure of our progenitor star, or a result of the
piston explosion mechanism.

\begin{figure}
  \includegraphics[width=0.45\textwidth]{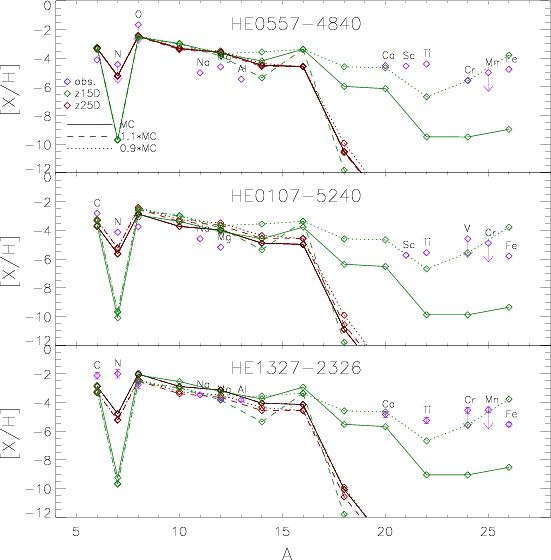}
  \caption{Model results matched to carbon and oxygen abundances
    found in individual HMP halo stars.  Model z15D, with less
    fallback, expels some iron peak elements.  Nitrogen is always
    under-produced, which is not a result of insufficient mixing
    during the supernova explosion but rather the presupernova
    evolution of these models.  The solid line is the yield
    calculated using the remnant masses from \cite{Zhang:2008}.  The
    dotted and dashed lines correspond to the amount of infall found
    in \cite{Zhang:2008} decreased or increased by $10\%$, respectively.
    The sensitivity of the yields to fallback is especially apparent
    for Model z15D, the model with the least mixing.\label{HMPcomp}}
\end{figure}

Because very little mixing occurs in the layers interior to the mass
cut of these zero metallicity stars, their yields were very sensitive
to the amount of fallback that occurs.  Figure \ref{HMPcomp} shows
this sensitivity. When the mass of infalling material was increased by
10\% of the fiducial value taken from \citet{Zhang:2008}, almost none of
the elements interior to the C-O shell escape.  When the mass of
infalling material was decreased by 10\% of this fiducial value, more
of the elements interior to oxygen escape and go on to enrich the
surrounding gas. Figure \ref{HMPcomp} shows a comparison of the yields
of our zero-metallicity stars with observations of the three most
metal-poor stars yet found in the halo of the Milky Way.  [X/H] values
for all elements were obtained by diluting the yields of our model
stars with enough pristine gas to fit observations of [C/H] and [O/H]
to individual halo stars.  The dotted and dashed lines show the yields
resulting from increasing or decreasing the amount of fallback by 10\%
of its fiducial value, respectively.  Model z15D is more sensitive
than Model z25D to decreasing the mass cut.  The model yields appear
much closer to fitting the values of elements heavier than oxygen
observed in the metal-poor halo stars.

The yields of our Pop III core collapse supernovae reproduce the
extreme overabundance of [C,O/Fe] to the point where Fe is
underproduced even when compared to the already low observations of
[C,O/Fe] in HMP stars.  Carbon is slightly underproduced and nitrogen
is underproduced by 2-3 dex relative to oxygen, as is the iron peak.
The iron peak elements are underproduced because they do not mix
sufficiently with the lighter elements in the star and cannot escape
falling back onto the remnant at the center.  A reduction in the
infall mass brings z15D closer to reproducing observations.  Two
dimensional versions of the one dimensional simulations presented in
the higher energy explosions of \citet{Heger&Woosley:2008} and \citet{Nomoto:2007} might
eject more of the \Ni \ produced in the more energetic explosions.
Rotating models \citep{Hirschi:2008} create more primary nitrogen, which can
lead to an increase in the rate of CNO burning at the base of the
hydrogen shell, causing some models die as larger red supergiants
rather than a compact blue supergiants. In this case, Rayleigh-Taylor
mixing would play out in a similar way to the solar models presented
in this paper, and more iron would be ejected by these stars.  Recent
simulations \citep{Scheck:2004,Scheck:2006,Burrows:2007a,Burrows:2007b, Burrows:2007c} point out that
the supernova explosion mechanism is probably inherently
multidimensional and asymmetric.  Asymmetry in the explosion, whether
in the form of a jet or a perturbation described by Legendre
polynomials of order of $l=1$ or $l=2$, might also mix more of the
nickel core out of the star, bringing the models closer to reproducing
observations. \citet{Venn&Lambert:2008} have suggested that HMP stars may be
''chemically peculiar'' stars, in which low iron abundance is caused
by separation of gas and dust beyond the stellar surface, followed by
accretion of dust-depleted gas.  If this is the case--and the authors
note that a definitive answer requires additional information--the
stars' true metallicity is closer to [X/H] $\approx -2$ rather than
-5.

\subsection{Visibility}
\lSect{visibility}

The supernova light curve is affected by the amount of $^{56}$Ni in
the center of the star that falls back onto the black hole at the
center of the explosion.  The models for the first supernovae
presented in this work are intrinsically dimmer than corresponding
supernovae arising from stars of solar composition provided they
explode with the same amount of energy. In our models of primordial
composition supernovae, all or nearly all of the \Ni \  synthesized in
the supernova falls back onto the remnant left behind at the center of
the explosion.  Energy from the radioactive decay of \Ni \  powers the
tail of core-collapse supernova light curves.  When the energy
released in its radioactive decay to $^{56}$Fe is no longer
observable, the supernova light curves will loose their radioactive
tails, making them briefer and dimmer than ordinary core-collapse
supernova light curves.

\section{CONCLUSIONS}
\lSect{conclusions}

The presupernova structure of a star is determined largely by its
initial mass and by the initial composition of the gas from which it
formed.  The symmetry and energy of the explosion, along with the
presupernova structure, influence where and to what extent
Rayleigh-Taylor instabilities will grow, as well as how much mass will
fall back onto the remnant at the center.  The non-rotating zero
metallicity models studied are far more compact than solar-composition
models of the same mass, in part because CNO burning proceeds at
higher temperatures and densities.  CNO burning is responsible for
energy production during the main sequence for all stars at the masses
studied here, but in metal poor stars CNO burning proceeds at higher
temperatures and densities. For zero-metalicity stars, the star must
first contract to a temperature of 10$^8$ K, hot enough to initiate
helium burning.  This helium burning produces a small amount of
carbon, which is enough to act as a catalyst to enable hot CNO burning
to proceed. In addition, non-rotating stars with a metallicity Z below
10$^{-3}$ will never reach the red giant branch, since they end helium
burning with effective temperatures above 10$^4$.  Below this
temperature, the opacity is large enough that the star will expand
toward the red giant branch. The more compact structure of these stars
causes their reverse shocks to propagate more quickly to the origin
than those in solar stars.  Larger remnants are left behind in the
more compact stars because the rate at which mass accretes onto the
stellar remnant is higher, as predicted by \citet{Chevalier:1989} and shown in
the 1D simulations of \citet{Zhang:2008}.

The time scale over which the Rayleigh-Taylor instabilities can
develop is also set by the reverse shock.  For the case of the
compact primordial composition progenitors modeled here, the
Rayleigh-Taylor instabilities have little time to develop.  This means
that a smaller portion of the isotopic layers of the star will be
mixed.  The Rayleigh-Taylor instabilities do not have time to become
fully nonlinear in our simulations, so the scale of the instability as
well as the degree of mixing is set by the scale of the initial seed
perturbations.  In the case of the solar-composition progenitor
models, the Rayleigh-Taylor instability became fully nonlinear and the
size and shape of the initial perturbation was no longer apparent at
late times.  A smaller region of the primordial-composition stars is
unstable, compared to solar-composition stars, which also
contributes to the reduced mixing we see in our zero-metal models.
The small amount of mixing experienced by zero-metallicity stars means
that their yields are very sensitive to the amount of fallback, which
may depend somewhat on the time the models were mapped from
\code{KEPLER} to \code{FLASH}.

Unlike the recent simulations of Rayleigh-Taylor mixing in compact
blue supergiants by \citet{Kifonidis:2003,Kifonidis:2006}, we do not see mixing at the
silicon-oxygen shell interface.  The absence of instability at that
interface, as well as at the nickel-silicon shell interface in zD
stars is robust. Mixing was not observed even after mapping the models
to \code{FLASH} at earlier times, running the models at significantly
higher resolution, and adding large perturbations of 10$\%$ to the
velocity.  Differences between the progenitor models used in
\citet{Kifonidis:2003,Kifonidis:2006} and our progenitor models, or structural
differences arising from their early modeling and different explosion
mechanism may be responsible.  The degree of mixing we observe in our
models is similar to that of earlier studies of compact blue
supergiant progenitor models for SN 1987A.  The degree of mixing and
final velocities of isotopes we observe in our solar composition
models is comparable to that found in \citet{Herant&Woosley:1994}. Material from the
silicon and nickel layers is mixed out into the outer layers of the
solar metallicity stars, and these isotopes are mixed to velocities of
$\sim 1-2\times 10^3$ km s$^{-1}$, velocities comparable to those
found in \citet{Herant&Woosley:1994}.

The zero-metallicity models show a dramatic overproduction in carbon
and oxygen relative to iron.  This is the trend observed in the most
metal-poor stars in the universe, but the degree of mixing in the
present work is so small that not enough iron escapes to reproduce the
already large [C+O/Fe] values observed in these stars.  Decreasing the
amount of fallback onto the remnant at the center through a more
energetic explosion might succeed in driving more iron-peak isotopes
out of the star, but would not result in a higher amount of nitrogen.
Our models also cannot reproduce the [C/N] or [O/N] ratios observed in
HMP stars.  Zero-metallicity stars with significant rotation might
produce both more primary nitrogen as well as more iron.  Rotating
models experience more shear mixing, which can drag primary carbon and
oxygen from the \He \ burning convective core to the \Hy \ burning
shell, where it becomes nitrogen, increasing the amount of burning in
the \Hy \ shell and making that shell convective.  The zero-metallicity
stars may then die as a red, rather than blue, supergiant
\citep{Hirschi:2008}.  If this is the case, then rotating zero-metallicity
models might look very much like the solar-metallicity models
presented in this paper, which would more closely reproduce abundances
observed in HMP stars.  A higher SN explosion energy might be another
way to expel more iron from these stars.

Future work will include modeling primordial composition models with
some degree of rotation, as recent work \citep{Hirschi:2008} indicates that
they should have more primary nitrogen, as well as less compact
structure and a larger \He \ core.  Modeling rotating stars may result
in drastically different yields for our zero metallicity stars,
allowing more iron to escape and increasing the [C/N] ratio closer to
what is observed in HMP stars.  We will explore higher explosion
energies and asymmetric explosions and how they affect the mount of
mixing.

\acknowledgments

The authors would like to thank Mike Zingale and Bruce Fryxell for
assistance with the \code{FLASH} code, and Brian O'Shea and Gabriel
Rockefeller for helpful comments on early drafts of this MS.  The
software used in this work was in part developed by the DOE-supported
ASC / Alliance Center for Astrophysical Thermonuclear Flashes at the
University of Chicago.  This research has been supported by the NASA
Theory Program NNG05GG08G and the DOE Program for Scientific Discovery
through Advanced Computing (SciDAC; grants DOE-FC02-01ER41176 and
DOE-FC02-06ER41438).  At LANL, Heger and Joggerst performed this work
under the auspices of the National Nuclear Security Administration of
the U.S.\ Department of Energy at Los Alamos National Laboratory under
Contract No.\ DE-AC52-06NA25396.

\bibliography{astro_ph}

\end{document}